\documentclass[a4paper,fleqn,usenatbib]{mnras}

\usepackage{newtxtext,newtxmath}
\usepackage[T1]{fontenc}
\usepackage{ae,aecompl}
\usepackage{graphicx}
\usepackage{amsmath}
\usepackage{amssymb}

\title[Environmental BAO in BOSS-CMASS]{The environmental dependence
  of the baryon acoustic peak in the Baryon Oscillation Spectroscopic
  Survey CMASS sample}

\author[Blake et al.]{Chris Blake,$^1$\thanks{E-mail:
    cblake@swin.edu.au} Ixandra Achitouv$^{1,2}$, Angela Burden$^3$
  and Yann Rasera$^2$ \\ $^1$ Centre for Astrophysics \&
  Supercomputing, Swinburne University of Technology, P.O.\ Box 218,
  Hawthorn, VIC 3122, Australia \\ $^2$ LUTH, UMR 8102 CNRS,
  Observatoire de Paris, PSL Research University, Universit\'{e} Paris
  Diderot, 92190 Meudon, France \\ $^3$ Department of Physics, Yale
  University, New Haven, CT 06511, USA}

\date{Accepted XXX. Received YYY; in original form ZZZ}

\pubyear{2018}

\begin{document}
\label{firstpage}
\pagerange{\pageref{firstpage}--\pageref{lastpage}}
\maketitle

\begin{abstract}
  The environmental dependence of galaxy clustering encodes
  information about the physical processes governing the growth of
  cosmic structure.  We analyze the baryon acoustic peak as a function
  of environment in the galaxy correlation function of the Baryon
  Oscillation Spectroscopic Survey CMASS sample.  Dividing the sample
  into three subsets by smoothed local overdensity, we detect acoustic
  peaks in the six separate auto-correlation and cross-correlation
  functions of the sub-samples.  Fitting models to these correlation
  functions, calibrated by mock galaxy and dark matter catalogues, we
  find that the inferred distance scale is independent of environment,
  and consistent with the result of analyzing the combined sample.
  The shape of the baryon acoustic feature, and the accuracy of
  density-field reconstruction in the Zeldovich approximation, varies
  with environment.  By up-weighting underdense regions and
  down-weighting overdense regions in their contribution to the
  full-sample correlation function, by up to $50\%$, we achieve a
  fractional improvement of a few per cent in the precision of baryon
  acoustic oscillation fits to the CMASS data and mock catalogues: the
  scatter in the preferred-scale fits to the ensemble of mocks
  improves from $1.45\%$ to $1.34\%$ (pre-reconstruction) and $1.03\%$
  to $1.00\%$ (post-reconstruction).  These results are consistent
  with the notion that the acoustic peak is sharper in underdense
  environments.
\end{abstract}


\section{Introduction}
\label{secintro}

Baryon acoustic oscillations (BAO) are a feature of the 2-point
clustering pattern of galaxies which encodes a preferred co-moving
separation of order $100 \, h^{-1}$Mpc -- the sound horizon at the
baryon drag epoch.  The feature manifests as a small but discernible
peak in the galaxy correlation function, and as a series of harmonics
in the galaxy power spectrum.

BAO in large-scale structure have emerged as an important cosmological
probe because this preferred length scale, calibrated by
early-Universe physics established by the Cosmic Microwave Background,
may be used as a standard ruler to map out the cosmic distance scale
and expansion rate as a function of redshift
\citep{Eisenstein98b,Blake03,Seo03,Eisenstein05}.  Recent large-scale
structure surveys have used BAO to report distance-scale measurements
for redshifts $z < 2.3$ with accuracies in the range $1-5\%$
\citep{Blake11,Beutler11,Padmanabhan12,Kazin14,Anderson14,Ross15,Bautista17,Alam17,Ata18}.
These measurements are consistent with the distance-redshift relation
inferred from analysis of the Cosmic Microwave Background, in the
context of the $\Lambda$CDM model and FLRW metric \citep{Planck18}.

Although the baryon acoustic peak is a robust prediction of
early-Universe physics \citep{Eisenstein04}, its presence in the
late-time clustering pattern is modulated by non-linear effects such
as the growth of structure, redshift-space distortions and galaxy bias
\citep{Eisenstein07a,Smith08,Seo08,Crocce08,Matsubara08}.  These
modifications may be used either as a source of additional
cosmological information, or as a mechanism for enhancing the accuracy
of the standard ruler by approximately recovering the pristine
information from early times.

An important property of the late-time acoustic peak is a broadening
caused by the displacement of galaxies from their original locations
in the density field, produced by bulk-flow motions which trace the
contraction of overdense regions and expansion of voids
\citep{Eisenstein07a,Sherwin12,McCullagh13,Rasera14}.  These
displacements may be estimated in linear perturbation theory from the
density field traced by the galaxies, and partially retracted in order
to sharpen the peak profile, in a process known as density-field
reconstruction \citep{Eisenstein07b,Padmanabhan09}.  The statistical
properties of the galaxy displacements and the accuracy of the
reconstruction algorithm depend on local environment, which controls
the growth of structure and applicability of linear theory
\citep{Achitouv15}.

As a consequence of these effects, the profile of the baryon acoustic
peak depends on local environment \citep{Neyrinck18}.  In this paper
we will map out this environmental dependence using the largest
current galaxy large-scale structure dataset, the Baryon Oscillation
Spectroscopic Survey (BOSS) \citep{Dawson13,Reid16,Alam17}.  Our
motivation is two-fold.  First, if the acoustic peak shape depends on
environment, then using a fitting template matched to each environment
might result in an improved error in the distance-scale fit.  Second,
if the sharpness of the baryon acoustic peak is a function of
environment, or if density-field reconstruction is less accurate in
dense environments owing to the increased importance of non-linear
effects, then enhancing the weight of low-density environments might
also improve the fit \citep{Achitouv15}.

A number of previous studies have explored the behaviour of the baryon
acoustic peak in the context of environment, with different emphasis
and aims.  \citet{Neyrinck18} found that the location of the acoustic
peak in the correlation function of N-body dark matter simulations was
shifted as a function of environment by a few per cent, due to the
contraction of overdense regions and expansion of underdense regions.
This effect was investigated using the Luminous Red Galaxy dataset of
the Sloan Digital Sky Survey by \citet{Roukema15}, who reported that
the acoustic peak is compressed by a similar factor for galaxy pairs
spanning supercluster regions.  \citet{Kitaura16} detected the
acoustic peak in the correlation function of a sample of voids
constructed from the BOSS dataset, and \citet{Zhao18} explored the
additional distance-scale information that resulted when these void
measurements were combined with galaxy clustering.  Although we focus
on correlation function analysis in this paper, we note that
clustering statistics sensitive to environment may be constructed in
several different ways, such as the sliced correlation function
\citep{Neyrinck18}, the density-marked correlation function
\citep{White16} and as position-dependent clustering sensitive to
higher-order correlations \citep{Chiang14}.  Finally,
density-dependent effects may be linked to a rich set of theoretical
phenomenology, such as screening mechanisms in modified gravity
scenarios \citep{Falck15} or the imprint of an inhomogeneous
cosmological metric \citep{Roukema16}.

This paper is structured as follows.  In Section \ref{secdata} we
present the data and mock catalogues utilized in our analysis, our
definitions of local environment, and our procedures for estimating
the galaxy correlation function within environmental slices and
weighting pair counts as a function of overdensity.  In Section
\ref{secmodel} we describe our approach for fitting BAO models as a
function of environment, and in Section \ref{secresults} we summarize
the results of our distance-scale fits to the environmental and
weighted correlation functions.  We present our conclusions in Section
\ref{secconc}.

\section{Measurements}
\label{secdata}

\subsection{Data and mock catalogues}

Our study is based on the final data release (DR12) of the Baryon
Oscillation Spectroscopic Survey \citep{Dawson13,Reid16,Alam17}.  In
particular we analyzed the largest component of BOSS, the CMASS
Luminous Red Galaxy sample, which was selected by optical colour and
magnitude cuts to form an approximately stellar-mass limited sample of
almost 1 million massive galaxies in the redshift range $0.43 < z <
0.7$.  The sample is divided into two contiguous regions of sky,
covering the Northern Galactic Cap (NGC) and Southern Galactic Cap
(SGC), which together span an effective volume of $5.1$ Gpc$^3$ for
cosmological purposes \citep{Reid16}, with an effective redshift
$z_{\rm eff} = 0.57$ which we adopted in this analysis.

CMASS galaxies are assigned a combination of weights for clustering
analysis \citep{Reid16}: observational systematics weights, which
correct for non-cosmological density fluctuations induced by varying
stellar density and seeing, weights which compensate for missing
objects due to fibre collisions or redshift failures, and FKP weights
\citep{Feldman94} which optimally balance the contribution of sample
variance and Poisson noise to clustering measurements.  We applied all
of these weights in our correlation function analysis.  We also
utilized the accompanying CMASS random catalogues, which are around 50
times larger than the galaxy dataset, and constructed to match its
redshift distribution and angular coverage.

Mock galaxy catalogues are an integral part of clustering studies,
allowing us to test model-fitting procedures on a simulated dataset
with known input cosmology, and determine covariance matrices from the
ensemble of mocks.  Our analysis utilized 600 Quick Particle Mesh
(QPM) mock galaxy catalogues \citep{White14} accompanying the CMASS
dataset.  These mock catalogues are built from low-resolution
particle-mesh simulations, from which sub-samples of particles are
drawn with properties approximately matching the distribution and
clustering statistics of dark matter halos.  These halo tracers are
populated with a halo occupation distribution and sub-sampled to match
the selection function and clustering of the CMASS sample (see
\citet{Alam17} for a summary of the mock catalogues generated for BOSS
clustering analyses).

As discussed in Section \ref{secintro}, galaxies are displaced from
their original positions in the density field by bulk motions induced
by the growth of cosmic structure \citep{Eisenstein07b}; these motions
broaden and dilute the baryon acoustic feature.  We computed the
displacement field in the Zeldovich approximation of the data and mock
catalogues, and corresponding random samples, using the Fourier-space
algorithm introduced by \citet{Burden15}.  We used this displacement
field to retract objects to their near-original positions and hence
sharpen the acoustic peak.  In the following sections we will present
results before and after this density-field reconstruction procedure.

We constructed one of our models for the dependence of the baryon
acoustic peak shape on environment using matched N-body dark matter
simulations evolved from two sets of initial conditions: a standard
fiducial power spectrum containing baryon acoustic oscillations, and a
``no-wiggles'' power spectrum which closely follows the smooth power
spectrum shape of the first simulation, removing the oscillations
\citep{Eisenstein98a}.  Through comparison of the resulting clustering
patterns as a function of environment, the effects on the baryon
acoustic peak may be distinguished from other clustering properties,
and the difference used to construct models.  These simulation
datasets were specifically generated to investigate BAO in
\citet{Rasera14}.  They cover a volume of $388$ Gpc$^3$ using $8.6$
billion particles and are part of the Dark Energy Universe Simulation
(DEUS) suite \citep{Alimi12,Rasera14}.

Different fiducial cosmological models, summarized in Table
\ref{tabcosmo}, were used to map the angular positions and redshifts
of CMASS galaxies into co-moving co-ordinates, and to construct the
mock catalogues discussed above.  Given that the theoretical sound
horizon scale depends on the fiducial matter and baryon densities, and
its observed position in the correlation function is distorted by
Alcock-Paczynski effects in a trial cosmology, these differences are
important to track.  We analyzed the clustering of the CMASS sample
and accompanying QPM mock catalogues, and performed density-field
reconstruction, using the BOSS DR12 fiducial cosmology \citep{Alam17}
listed in the second column of Table \ref{tabcosmo}.  However, the QPM
mock catalogues and DEUS simulations were constructed from fiducial
power spectra of different cosmological models, with varying
sound-horizon scales, as summarized in the third and fourth columns of
Table \ref{tabcosmo}.  These differences were fully accounted in our
BAO-fitting process, as explained in Section \ref{secmodel}.

\begin{table}
  \caption{The fiducial cosmological models relevant to our analysis.
    We measured correlation functions from the CMASS data and QPM
    mocks using the ``BOSS DR12'' fiducial cosmology listed in the
    second column.  This model differs from the cosmology used to
    construct the initial power spectrum of the QPM mocks (third
    column) and the DEUS wiggles and no-wiggles dark matter
    simulations (fourth column).  The sound horizon at baryon drag,
    $r_s(z_d)$, is computed for each cosmological model from CAMB; our
    value of $r_s(z_d)$ for the BOSS DR12 fiducial cosmology differs
    slightly from that quoted in \citet{Alam17} because we assume
    $\Omega_\nu = 0$.  The final row of the table lists the
    volume-weighted distance $D_V$ for each cosmological model,
    evaluated at the effective redshift $z=0.57$.}
\begin{center}
\begin{tabular}{|c|c|c|c|}
\hline
Parameter & BOSS DR12 & QPM mock & DEUS \\
\hline
$\Omega_{\rm m}$ & $0.31$ & $0.29$ & $0.2573$ \\
$\Omega_{\rm b}$ & $0.0481$ & $0.0459$ & $0.0436$ \\
$h$ & $0.676$ & $0.7$ & $0.72$ \\
$\sigma_8$ & $0.8$ & $0.8$ & $0.801$ \\
$n_s$ & $0.97$ & $0.97$ & $0.963$ \\
\hline
$r_s(z_d)$ [Mpc] & $147.62$ & $147.1$ & $149.37$ \\
\hline
$D_V(z=0.57)$ [Mpc] & $2059.6$ & $2009.5$ & $1988.2$ \\
\hline
\end{tabular}
\end{center}
\label{tabcosmo}
\end{table}

\subsection{Defining and weighting the environments}
\label{secenv}

In this study we investigate the dependence of the large-scale
correlation function of the CMASS galaxy dataset on local environment.
We defined environment as the local overdensity of the galaxy sample,
$\delta(\vec{x})$ at position $\vec{x}$, smoothed using a Gaussian
filter with standard deviation $15 \, h^{-1}$Mpc,
\begin{equation}
\delta(\vec{x}) = \frac{N_R}{N_D} \, \frac{\rm{Sm}(D)}{\rm{Sm}(R)} - 1 ,
\end{equation}
where ${\rm Sm}()$ represents the application of Gaussian smoothing to
the data catalogue ($D$) containing $N_D$ galaxies, or the random
catalogue ($R$) of $N_R$ points.

The adopted smoothing scale of $15 \, h^{-1}$Mpc and resulting
overdensity field are identical to those used in the density-field
reconstruction of the sample \citep{Burden15,Alam17}.  The dependence
of the performance of reconstruction on smoothing scale has been
investigated by \citet{Vargas17} and \citet{Achitouv15}.  For the
CMASS sample, the recovered isotropic baryon acoustic scale is
insensitive to the choice of smoothing scale within the range $5$ to
$15 \, h^{-1}$Mpc, although smoothing scales closer to $5 \,
h^{-1}$Mpc may yield better performance for anisotropic fits
\citep{Vargas17} or for a dataset with higher number density
\citep{Achitouv15}.

\begin{figure}
\includegraphics[width=\columnwidth]{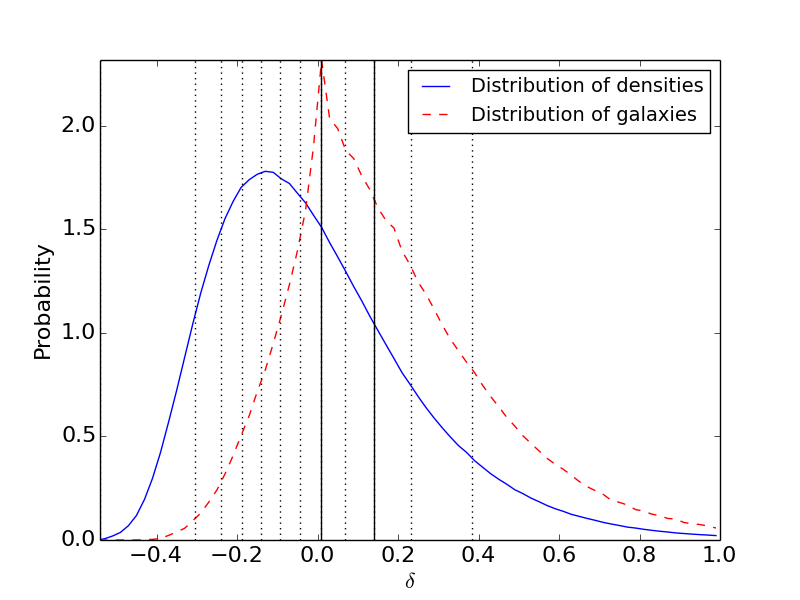}
\caption{The distribution of overdensity values $\delta$ within the
  CMASS-North region grid as traced by random points (solid blue
  histogram), and as traced by CMASS-North galaxies (dashed red
  histogram, whose distribution is naturally skewed towards higher
  values of $\delta$).  The vertical dotted lines separate the 12
  narrow density bins in which we initially measured the galaxy pair
  counts, and the vertical solid lines separate the 3 environmental
  slices we used to construct the final galaxy auto-correlation and
  cross-correlation functions.}
\label{figdelhist}
\end{figure}

Following the construction of the smoothed density field, we assigned
values of $\delta$ to the data and random points using cloud-in-cell
interpolation.  The frequency distribution of $\delta$ across the
survey volume (i.e.\ of the random points) is displayed as the blue
solid curve in Figure \ref{figdelhist}, together with the distribution
of overdensities assigned to the galaxies as the red dashed curve,
which is naturally skewed towards higher values of $\delta$, given
that more galaxies are located in overdense environments.

We defined $N_{\rm env} = 12$ narrow density bins by dividing the
survey such that each bin contains equal volume; these bin divisions
are displayed as the vertical dotted lines in Figure \ref{figdelhist}.
We ultimately grouped these environments into clustering measurements
in 3 density slices, such that the baryon acoustic peak could be
detected in each correlation function.  Initially measuring the pair
counts in narrow bins afforded us the flexibility to adjust the
groupings within the coarser bins, without needing to repeat the
original pair count measurements.

In addition to constraining any variation of the acoustic peak scale
with environment, we also wished to investigate whether acoustic peak
measurements may be improved by weighting pair counts as a function of
environment.  Following \citet{Achitouv15}, we adopted a simple
weighting scheme where the weights $w_i$ of the $N_{\rm env} = 12$
environments ($1 \le i \le N_{\rm env}$) varied uniformly as,
\begin{equation}
w_i = 1 + x \left[ \frac{2 \, (i-1)}{N_{\rm env}-1} - 1 \right] ,
\label{eqweight}
\end{equation}
in terms of a single parameter $x$ constrained to lie in the range $-1
\le x \le 1$.  Hence, for a given choice of $x$, $w_i$ varied
uniformly from $1-x$ to $1+x$, such that negative values of $x$ have
the effect of upweighting underdense environments with respect to
overdense environments.

\subsection{Estimating the environmental correlation function}
\label{seccorrest}

The difference in the distribution of environments in which data and
random points reside (Figure \ref{figdelhist}) complicates the
estimation of the correlation function within an environment slice: if
the redshift distributions of the parent data and random samples are
consistent when averaged over all environments, then the redshift
distributions of the corresponding environmental sub-samples will
disagree.  Therefore, the parent random catalogue requires a
redshift-dependent sub-sampling or weighting in order to match the
galaxy redshift distribution in a density slice, and provide
appropriate pair counts for estimating the environmental correlation
function.

We implemented this procedure by measuring pair counts in bins of
environment, separation and redshift (co-moving distance in the
fiducial cosmology), and then scaling the resulting counts as a
function of redshift in the manner described below.  We used the 12
environment bins defined in Section \ref{secenv}, 36 separation bins
of width $5 \, h^{-1}$ Mpc between $0$ and $180 \, h^{-1}$ Mpc, and 14
bins of co-moving distance of width $50 \, h^{-1}$ Mpc between $1100$
and $1800 \, h^{-1}$ Mpc.  We hence measured the data-data pair counts
$dd_{ijkl}(s)$, data-random pair counts $dr_{ijkl}(s)$ and
random-random pair counts $rr_{ijkl}(s)$ in separation bins $s$,
between bins of environment (denoted by indices $i$ and $j$ for the
two catalogues entering the pair count) and redshift (denoted by
indices $k$ and $l$ for the two catalogues).  We measured pair counts
using the {\tt corrfunc} software \citep{Corrfunc}.

We now outline how we weighted these pair counts when measuring the
galaxy correlation function in a density slice $I$, consisting of some
combination of the environments $i$, and the cross-correlation
function of galaxies in density slices $I$ and $J$.  We also allowed
for weighting the contribution of each environment $i$ to the pair
count by $w_i$, as defined in Equation \ref{eqweight}.  In the
following, we write the number of data and random points in
environment $i$ and redshift bin $k$ as $n_{D,ik}$ and $n_{R,ik}$,
respectively.

The weighted number of data and random sources in each redshift bin
$k$, summed over the combination of environments $i$ for which we wish
to measure the correlation function, is then,
\begin{equation}
  \begin{split}
    N_{D,Ik} &= \sum_i w_i \, n_{D,ik} , \\
    N_{R,Ik} &= \sum_i w_i \, n_{R,ik} ,
  \end{split}
\end{equation}
with total numbers of objects in the density slice,
\begin{equation}
  \begin{split}
    N_{D,I} &= \sum_i \sum_k w_i \, n_{D,ik} , \\
    N_{R,I} &= \sum_i \sum_k w_i \, n_{R,ik} .
  \end{split}
\end{equation}
We defined weighted pair counts between density slices $I$ and $J$,
summed over the corresponding environmental sub-samples $i$ and $j$
and redshift bins $k$ and $l$, as,
\begin{equation}
  \begin{split}
    DD_{IJ}(s) &= \sum_i \sum_j \sum_k \sum_l w_i \, w_j \, dd_{ijkl}(s) , \\
    DR_{IJ}(s) &= \sum_i \sum_j \sum_k \sum_l w_i \, w_j \, \frac{N_{D,l}}{N_{R,l}} \, dr_{ijkl}(s) , \\
    RD_{IJ}(s) &= \sum_i \sum_j \sum_k \sum_l w_i \, w_j \, \frac{N_{D,k}}{N_{R,k}} \, dr_{jilk}(s) , \\
    RR_{IJ}(s) &= \sum_i \sum_j \sum_k \sum_l w_i \, w_j \, \frac{N_{D,k}}{N_{R,k}} \, \frac{N_{D,l}}{N_{R,l}} \, rr_{ijkl}(s) .
  \end{split}
\end{equation}
Finally, we estimated the correlation function between density slices
$I$ and $J$ using the estimator \citep{Landy93},
\begin{equation}
 \xi_{IJ}(s) = \frac{DD_{IJ}(s) - DR_{IJ}(s) - RD_{IJ}(s) +
   RR_{IJ}(s)}{RR_{IJ}(s)} .
\end{equation}
When estimating the post-reconstruction correlation function, we also
measured the pair counts of the random catalogue shifted by the
displacement field, denoted by $S$, such that \citep{Padmanabhan12},
\begin{equation}
 \xi_{IJ}(s) = \frac{DD_{IJ}(s) - DS_{IJ}(s) - SD_{IJ}(s) +
   SS_{IJ}(s)}{RR_{IJ}(s)} .
\end{equation}

\subsection{Correlation function measurements}

We used the approach described in Section \ref{seccorrest} to measure
the environmental auto- and cross-correlation functions $\xi_{IJ}$ of
the BOSS DR12 CMASS sample, the QPM mock catalogues, and the DEUS
wiggles and no-wiggles dark matter simulations.  We performed these
measurements in three density slices $I,J = \{ 1,2,3 \}$ where, if the
survey volume elements are ranked in order of increasing local
overdensity, the three slices span volumes in the ratio 7:2:3 (and are
constructed by summing pair counts for corresponding numbers of the
original 12 narrow environment bins).  The three density slices
correspond to overdensity ranges $-1 < \delta < 0.01$, $0.01 < \delta
< 0.14$ and $\delta > 0.14$ for $I,J = \{ 1,2,3 \}$, respectively.  We
chose these three density slices ensuring that we detected the baryon
acoustic peak in each individual auto- and cross-correlation function.
Given that $\xi_{JI} = \xi_{IJ}$, our three density slices yielded six
correlation functions, $\{ \xi_{11}, \xi_{12}, \xi_{13}, \xi_{22},
\xi_{23}, \xi_{33} \}$.  We also measured the total correlation
functions $\xi_{\rm tot}$ combining all environments of the different
samples, and the weighted correlation functions for different values
of the parameter $x$ defined in Equation \ref{eqweight}, spaced by
$\Delta x = 0.2$ in the range $-1 < x < 1$.

By determining the same correlation functions for the ensemble of mock
catalogues, we built a covariance matrix spanning the $\xi_{IJ}$
measurements for different density slices and scales.  For the BOSS
DR12 data and mock catalogues, we measured separate correlation
functions for the NGC and SGC survey regions, and combined these
measurements using inverse-variance weighting based on the correlation
function errors determined from the mocks.

\begin{figure}
\includegraphics[width=\columnwidth]{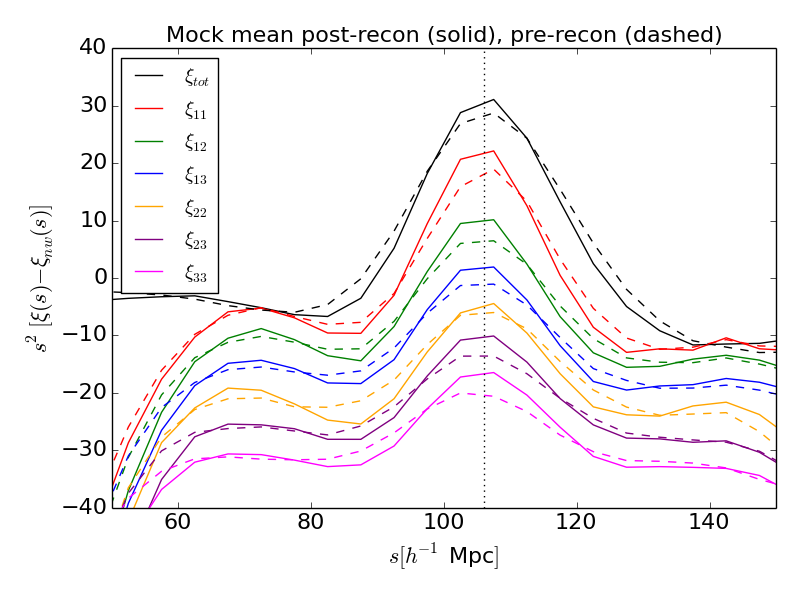}
\caption{The mean auto-correlation and cross-correlation functions
  $\xi_{IJ}(s)$ of the CMASS QPM mock galaxy catalogues in three
  environmental slices defined by increasing local density, compared
  to the total correlation function $\xi_{\rm tot}$.  In order to
  highlight the behaviour of the baryon acoustic peak, we subtracted a
  ``no-wiggles'' correlation function model from each measurement and
  scaled the resulting difference by $s^2$.  The solid and dashed lines
  display the correlation functions after and before reconstruction,
  respectively.  Successive correlation functions have been offset in
  the $y$-direction by $\Delta y = -5$, for clarity of presentation.
  The acoustic peak shape varies with environment.  The vertical
  dotted line, plotted at the approximate correlation function peak,
  indicates that the preferred scale does not vary significantly with
  environment.}
\label{figbaoenv}
\end{figure}

Figure \ref{figbaoenv} displays the mean of the total and
environmental correlation functions of the QPM mock catalogues, before
and after density-field construction, as the dashed and solid lines,
respectively.  We present these measurements after subtracting the
smooth ``no-wiggles'' component of the correlation function model,
described in more detail in Section \ref{secmodel}, to facilitate
comparison of the acoustic peak shape.  We find that the peak shape
depends on density, with the environmental correlation functions
exhibiting more prominent negative ``wings'' on either side of the
peak than the total correlation function, and the amplitude of the
baryon acoustic feature increasing steadily towards underdense
environments, varying by a factor of around two.  Density-field
reconstruction sharpens the acoustic peaks as expected.  We consider
the distance-scale fits to the correlation functions in Section
\ref{secresults} below.

\begin{figure*}
\includegraphics[width=14cm]{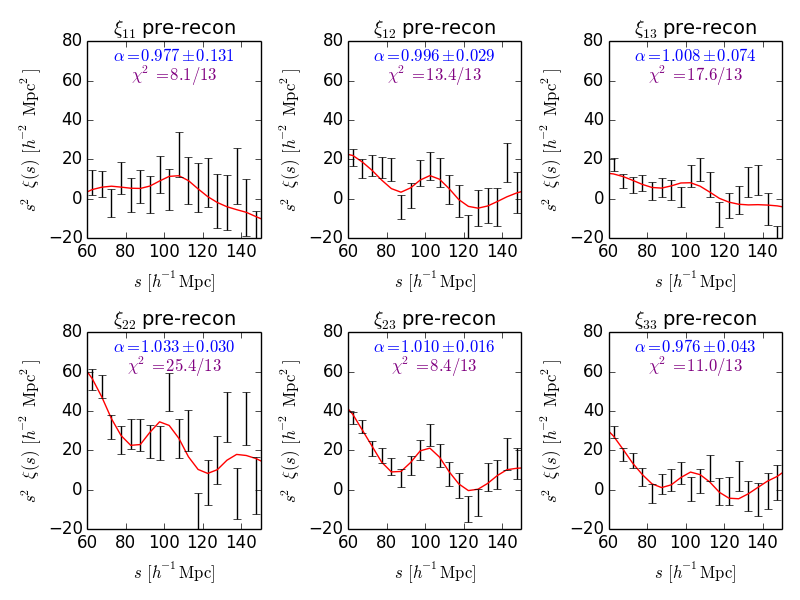}
  
\vspace{5mm}

\includegraphics[width=14cm]{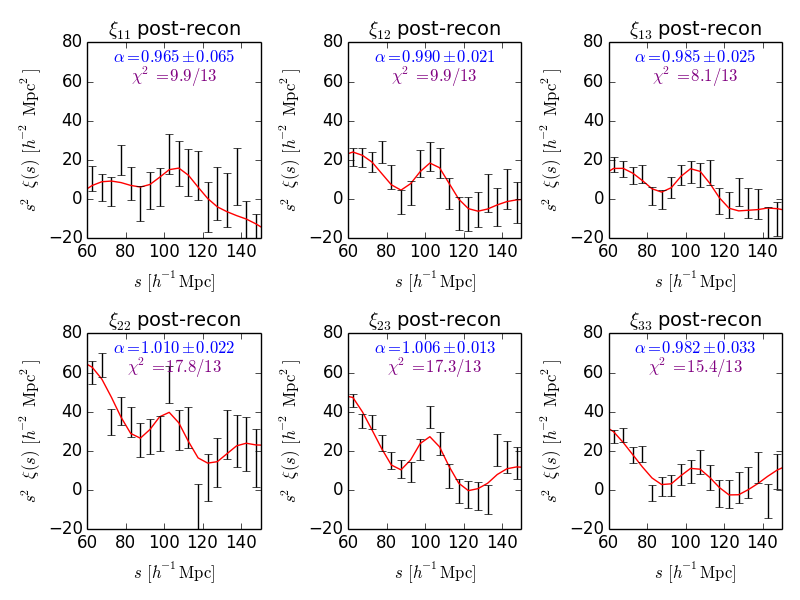}
\caption{Auto-correlation and cross-correlation function measurements
  $\xi_{IJ}$ of the CMASS DR12 galaxy dataset in three environmental
  slices defined by increasing local density.  The upper and lower
  sets of panels display results before and after reconstruction,
  respectively.  The solid line is the best-fitting model in each
  case, using a template built from the corresponding QPM mock mean
  correlation function.  We also show the $68\%$ confidence range of
  the posterior probability distribution of $\alpha$, and the $\chi^2$
  statistic of the best-fitting model and number of degrees of
  freedom.  Baryon acoustic peaks are detected in all six of the
  post-reconstruction environmental correlation functions.}
\label{figxienv}
\end{figure*}

Figure \ref{figxienv} presents the measurements of the auto- and
cross-correlation functions $\xi_{IJ}$ in density slices for the DR12
CMASS data sample, before reconstruction (upper panel) and after
reconstruction (lower panel).  The errors in the measurements are
determined from the diagonal elements of the covariance matrix built
from the mock catalogues.  We detect the baryon acoustic feature in
all six of the post-reconstruction correlation functions, and we
describe the fitted BAO models in the following section.

\section{BAO fitting}
\label{secmodel}

We considered three different models for the baryon acoustic peak in
the environmental and total galaxy correlation functions:
\begin{itemize}
\item A model constructed from a theoretical matter power spectrum,
\item A model built from the QPM mock galaxy correlation functions,
\item A model based on correlation functions of N-body dark matter
  simulations including and excluding baryon acoustic oscillations.
\end{itemize}

Each model describes distortions in a template correlation function
$\xi_{\rm template}(s)$ using 5 free parameters: a scale distortion
parameter $\alpha$, a correlation function amplitude $B^2$, and three
nuisance parameters $(A_0, A_1, A_2)$ of a polynomial which
marginalizes over the smooth shape of the correlation function,
ensuring it contributes no information to the scale fits
\citep{Anderson14}.  The form of these models is,
\begin{equation}
  \xi(s) = B^2 \xi_{\rm template}(\alpha s) + A_0 + A_1/s + A_2/s^2 ,
\label{eqxitemplate}
\end{equation}
where the construction of $\xi_{\rm template}$ is discussed below.  We
performed model fits using {\tt emcee} \citep{Foreman13}, adopting
uniform wide priors for the free parameters.  We fitted the
correlation function measurements in the range $60 < s < 150 \,
h^{-1}$Mpc, testing that this choice resulted in fits to the data and
mocks with acceptable values of the $\chi^2$ statistic, and that our
results did not significantly depend on the chosen fitting range.

In our first version of the model, we constructed $\xi_{\rm
  template}(s)$ following standard methods for fitting the baryon
acoustic peak in the angle-averaged total galaxy correlation function
\citep{Anderson14}, using a model power spectrum $P_{\rm mod}(k)$
generated assuming a fiducial cosmology,
\begin{equation}
  \xi_{\rm template}(s) = \frac{1}{2\pi^2} \int dk \, k^2 \, P_{\rm
    mod}(k) \, j_0(ks) \, e^{-k^2 s_{\rm damp}^2} ,
\label{eqximod}
\end{equation}
where $j_0(x)$ is the zeroth-order spherical Bessel function and
$s_{\rm damp} = 1 \, h^{-1}$Mpc.  The model power spectrum in Equation
\ref{eqximod} is calculated as,
\begin{equation}
P_{\rm mod}(k) = P_{\rm nw}(k) \left[ 1 + \left( \frac{P_{\rm
      lin}(k)}{P_{\rm nw}(k)} - 1 \right) e^{-\frac{1}{2} k^2
    \Sigma_{\rm nl}^2} \right] ,
\end{equation}
where $P_{\rm lin}(k)$ is the linear matter power spectrum computed
using the CAMB software \citep{Lewis00} in the fiducial cosmology,
$P_{\rm nw}(k)$ is a model created using the no-wiggles matter power
spectrum fitting formulae of \citet{Eisenstein98a}, and $\Sigma_{\rm
  nl}$ parameterizes the damping of the acoustic peak due to galaxy
displacements.  Following \citet{Anderson14}, we fixed $\Sigma_{\rm
  nl} = 8.3$ and $4.6 \, h^{-1}$Mpc for our fits to the
pre-reconstruction and post-reconstruction correlation functions,
respectively, noting that these choices (or indeed marginalizing over
$\Sigma_{\rm nl}$ as a free parameter) have no significant effect on
the results.

The linear power spectrum generated by CAMB may not be an appropriate
model for the clustering pattern in an environmental slice, if the
density-dependent power spectrum $P(k|\delta)$ has a different shape
to the total power spectrum $P(k)$ \citep{Chiang14}, and in Figure
\ref{figbaoenv} we indeed find that the acoustic peak shape varies
with environment.  We therefore considered two additional methods for
producing the template correlation function $\xi_{\rm template}(s)$
used in Equation \ref{eqxitemplate} to fit the baryon acoustic peak.

In our second model, we constructed the template using the mean
correlation functions of the QPM mocks, measured using the same
environmental binning as applied to the galaxy data.  The
environmental dependence of the correlation function shape is
therefore included in the fitted templates for this model.  We
retained free parameters for the correlation function amplitude and
polynomial terms, following Equation \ref{eqxitemplate}.

We built our third and final model for the acoustic peak from the DEUS
N-body dark matter simulations generated from two sets of initial
conditions: a linear CAMB power spectrum, and a no-wiggles power
spectrum with equivalent cosmological parameters.  We measured the
correlation functions of these two sets of dark matter particles,
$\xi_{\rm wig}(s)$ and $\xi_{\rm nw}(s)$, using the same environmental
binning as applied to the galaxy data, and constructed a model
correlation function from these measurements as,
\begin{equation}
  \xi(s) = B^2 \left[ \xi_{\rm wig}(\alpha s) - \xi_{\rm nw}(\alpha s)
    \right] + \xi_{\rm nw}(\alpha s) + A_0 + A_1/s + A_2/s^2 ,
\end{equation}
following the form proposed by \citet{Kitaura16}, tested on the
correlation function of minima of the density field.

Given that our correlation function templates are generated using a
range of model cosmologies with different corresponding standard ruler
scales, as summarized in Table \ref{tabcosmo}, we scaled the fitted
values of $\alpha$ such that they were all referenced to the fiducial
BOSS DR12 cosmology to enable a consistent comparison of results,
\begin{equation}
  D_V = \alpha \, D_{V,{\rm DR12-fid}} ,
\end{equation}
where $D_V$ is the volume-weighted distance measured by BAO in the
angle-averaged correlation function, defined in terms of the angular
diameter distance $D_A(z)$ and Hubble parameter $H(z)$ as,
\begin{equation}
D_V(z) = \left[ \frac{c \, z \, D_A(z)^2}{H(z)} \right]^{1/3} ,
\end{equation}
where $c$ is the speed of light and $D_{V,{\rm DR12-fid}}(z=0.57) =
2059.6$ Mpc.

Discussing this last point in more detail, suppose that we fit a
baryon acoustic peak with observed standard ruler scale $s_{\rm obs}$,
using a template correlation function with model standard ruler scale
$s_{\rm mod}$, which may differ from the true cosmological standard
ruler scale $s_{\rm true}$.  Suppose further that we measure the
correlation function of the data using a fiducial distance scale
$D_{V,{\rm fid}}$, which may differ from the true cosmological
distance scale $D_{V,{\rm true}}$.  In this case, the Alcock-Paczynski
distortion of the scale is given by $s_{\rm obs} = s_{\rm true} \,
(D_{V,{\rm fid}}/D_{V,{\rm true}})$, hence the expected best-fitting
value of $\alpha$ is,
\begin{equation}
\alpha_{\rm exp} = \frac{s_{\rm mod}}{s_{\rm obs}} = \frac{s_{\rm
    mod}}{s_{\rm true}} \, \frac{D_{V,{\rm true}}}{D_{V,{\rm fid}}} .
\end{equation}
If the template correlation function is calibrated as a function of
scale in units of $h^{-1}$Mpc, and if the values of the Hubble
parameter are $h_{\rm mod}$ and $h_{\rm fid}$ in the model and
fiducial cosmologies respectively, then we obtain,
\begin{equation}
\alpha_{\rm exp} = \frac{s_{\rm mod}[{\rm Mpc}]}{s_{\rm true}[{\rm
      Mpc}]} \, \frac{D_{V,{\rm true}}[{\rm Mpc}]}{D_{V,{\rm fid}}[{\rm
    Mpc}]} \, \frac{h_{\rm mod}}{h_{\rm fid}} .
\end{equation}
This relation motivates us to calibrate our fitted scale distortion
parameters to always refer to the BOSS DR12 fiducial cosmology, as
$\alpha_{\rm cal} = \alpha / \alpha_{\rm exp}$.  All values of
$\alpha$ quoted in the remainder of this paper are values of
$\alpha_{\rm cal}$.

In order to check the validity of our fitting procedures, we fitted
the QPM mock mean and DEUS correlation functions using all three
models described in this section, ensuring that we obtained results
consistent with $\alpha_{\rm cal} = 1$ after including the appropriate
calibration factors discussed above.

\section{Results}
\label{secresults}

\subsection{BAO fits to the total correlation function}

We initially fitted the three models described in Section
\ref{secmodel} to the total CMASS galaxy correlation function
$\xi_{\rm tot}$ with no environmental binning, before and after
density-field reconstruction.  Fitting the post-reconstruction
correlation function with the CAMB template we obtain $\alpha_{\rm
  cal} = 0.983 \pm 0.009$ (referenced to the BOSS DR12 fiducial
cosmology), and find that the best-fitting model is a good fit to the
data, with a $\chi^2$ statistic of $17.8$ for 13 degrees of freedom.
The other marginalized measurements of $\alpha_{\rm cal}$ are reported
in Table \ref{tabalphafits}.

\begin{table}
  \caption{BAO fits to the total correlation function $\xi_{\rm tot}$
    of the CMASS DR12 galaxy dataset, and to the auto-correlation and
    cross-correlation functions $\xi_{IJ}$, in three environmental
    slices defined by increasing local density.  Results are shown
    before and after density-field reconstruction, comparing models
    constructed from a CAMB power spectrum, the QPM mock mean, and the
    DEUS wiggles and no-wiggles dark matter simulations.  We report
    the $68\%$ confidence ranges of the posterior probability
    distribution for $\alpha_{\rm cal}$ (referenced to the BOSS DR12
    fiducial cosmology), together with the $\chi^2$ statistic of the
    best-fitting model and the number of degrees of freedom.}
\begin{center}
\begin{tabular}{|c|c|c|c|c|}
\hline
Recon & Data & Template & $\alpha_{\rm cal}$ & $\chi^2$/dof \\
\hline
Pre-recon & $\xi_{\rm tot}$ & CAMB & $0.997 \pm 0.013$ & $15.1/13$ \\
& & QPM mock & $0.992 \pm 0.014$ & $15.7/13$ \\
& & Wig/no-wig sim & $0.990 \pm 0.012$ & $13.9/13$ \\
\hline
Post-recon & $\xi_{\rm tot}$ & CAMB & $0.983 \pm 0.009$ & $17.8/13$ \\
& & QPM mock & $0.981 \pm 0.008$ & $16.6/13$ \\
& & Wig/no-wig sim & $0.984 \pm 0.009$ & $18.9/13$ \\
\hline
Pre-recon & $\xi_{11}$ & CAMB & $0.960 \pm 0.135$ & $7.9/13$ \\
& & QPM mock & $0.977 \pm 0.131$ & $8.1/13$ \\
& & Wig/no-wig sim & $0.957 \pm 0.125$ & $7.9/13$ \\
& $\xi_{12}$ & CAMB & $1.035 \pm 0.063$ & $16.4/13$ \\
& & QPM mock & $0.996 \pm 0.029$ & $13.4/13$ \\
& & Wig/no-wig sim & $0.974 \pm 0.026$ & $12.5/13$ \\
& $\xi_{13}$ & CAMB & $0.951 \pm 0.109$ & $17.2/13$ \\
& & QPM mock & $1.008 \pm 0.074$ & $17.6/13$ \\
& & Wig/no-wig sim & $0.986 \pm 0.055$ & $16.7/13$ \\
& $\xi_{22}$ & CAMB & $1.023 \pm 0.035$ & $28.2/13$ \\
& & QPM mock & $1.033 \pm 0.030$ & $25.4/13$ \\
& & Wig/no-wig sim & $1.016 \pm 0.029$ & $27.3/13$ \\
& $\xi_{23}$ & CAMB & $1.028 \pm 0.019$ & $11.6/13$ \\
& & QPM mock & $1.010 \pm 0.016$ & $8.4/13$ \\
& & Wig/no-wig sim & $0.994 \pm 0.014$ & $6.5/13$ \\
& $\xi_{33}$ & CAMB & $0.971 \pm 0.038$ & $9.1/13$ \\
& & QPM mock & $0.976 \pm 0.043$ & $11.0/13$ \\
& & Wig/no-wig sim & $0.965 \pm 0.032$ & $11.0/13$ \\
\hline
Post-recon & $\xi_{11}$ & CAMB & $0.959 \pm 0.081$ & $10.0/13$ \\
& & QPM mock & $0.965 \pm 0.065$ & $9.9/13$ \\
& & Wig/no-wig sim & $0.969 \pm 0.057$ & $9.4/13$ \\
& $\xi_{12}$ & CAMB & $1.006 \pm 0.039$ & $15.4/13$ \\
& & QPM mock & $0.990 \pm 0.021$ & $9.9/13$ \\
& & Wig/no-wig sim & $0.984 \pm 0.018$ & $7.4/13$ \\
& $\xi_{13}$ & CAMB & $0.975 \pm 0.033$ & $8.4/13$ \\
& & QPM mock & $0.985 \pm 0.025$ & $8.1/13$ \\
& & Wig/no-wig sim & $0.986 \pm 0.025$ & $7.5/13$ \\
& $\xi_{22}$ & CAMB & $1.010 \pm 0.035$ & $21.5/13$ \\
& & QPM mock & $1.010 \pm 0.022$ & $17.8/13$ \\
& & Wig/no-wig sim & $1.000 \pm 0.024$ & $18.9/13$ \\
& $\xi_{23}$ & CAMB & $1.013 \pm 0.015$ & $22.9/13$ \\
& & QPM mock & $1.006 \pm 0.013$ & $17.3/13$ \\
& & Wig/no-wig sim & $0.994 \pm 0.011$ & $12.8/13$ \\
& $\xi_{33}$ & CAMB & $0.979 \pm 0.046$ & $17.5/13$ \\
& & QPM mock & $0.982 \pm 0.033$ & $15.4/13$ \\
& & Wig/no-wig sim & $0.976 \pm 0.027$ & $14.5/13$ \\
\hline
\end{tabular}
\end{center}
\label{tabalphafits}
\end{table}

We find that the different BAO templates produce consistent
distance-scale fits to the whole-sample correlation function (after
correction for the varying model cosmologies as described above), and
that our results are consistent with those previously reported by the
BOSS collaboration.  For example, \citet{Cuesta16} reported a distance
measurement $D_V(z=0.57) = 2028 \pm 21$ Mpc using the
post-reconstruction isotropic CMASS correlation function of BOSS DR12,
compared to our measurement $D_V = 2025 \pm 19$ Mpc.

\subsection{Variation of BAO scale with environment}

We then fitted the auto-correlation and cross-correlation functions
$\xi_{IJ}$ of the CMASS dataset in the three environmental slices
defined by local density, before and after density-field
reconstruction, using covariance matrices constructed from
corresponding correlation functions of the QPM mocks.  We report the
marginalized measurements of $\alpha_{\rm cal}$ for all these cases in
Table \ref{tabalphafits}.

We can successfully detect and fit the baryon acoustic peak for each
of the six correlation functions, and the minimum values of the
$\chi^2$ statistic are consistent with the distribution expected from
the number of degrees of freedom if the data is drawn from the model
(the $\chi^2$ value of the fit to the pre-reconstruction $\xi_{22}$ is
high, but consistent with expected statistical fluctuations).  The
different models produce consistent values of $\alpha_{\rm cal}$.
Given that the CAMB power spectrum model may not be applicable to the
clustering of environmental slices as discussed in Section
\ref{secmodel}, we select the model constructed from the QPM mock mean
template as our fiducial choice in the following analysis, although
the other models produce consistent results.

\begin{figure*}
\includegraphics[width=15cm]{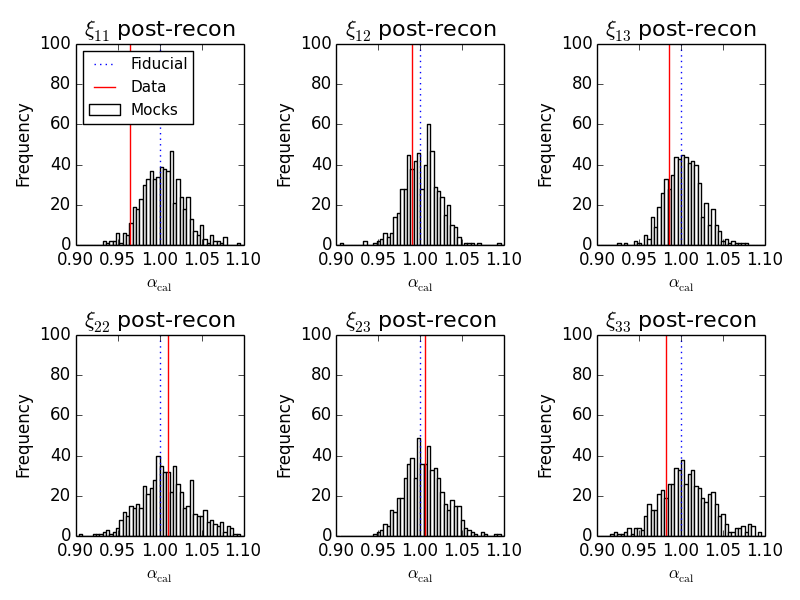}
\caption{Histogram of best-fitting $\alpha_{\rm cal}$ values fit to
  the post-reconstruction auto-correlation and cross-correlation
  functions $\xi_{IJ}$ of each of the 600 QPM mock galaxy catalogues,
  in three environmental slices defined by increasing local density.
  The QPM mock mean template is used to construct the BAO model,
  although the results are not sensitive to this choice.  The vertical
  solid red line illustrates the best-fit to the galaxy data in each
  case, and the vertical dotted blue line represents the expected
  fiducial value $\alpha_{\rm cal} = 1$ for fits to the mock
  catalogues using templates built from the mock mean.}
\label{figalphahist}
\end{figure*}

We performed corresponding fits to the environmental correlation
functions of each of the 600 individual QPM mock catalogues.  Figure
\ref{figalphahist} displays the distribution of best-fitting
$\alpha_{\rm cal}$ values across the mocks for each
post-reconstruction environmental correlation function $\xi_{IJ}$,
superimposing the values obtained from the data, which are consistent
with the mock distribution.

\begin{figure}
\includegraphics[width=\columnwidth]{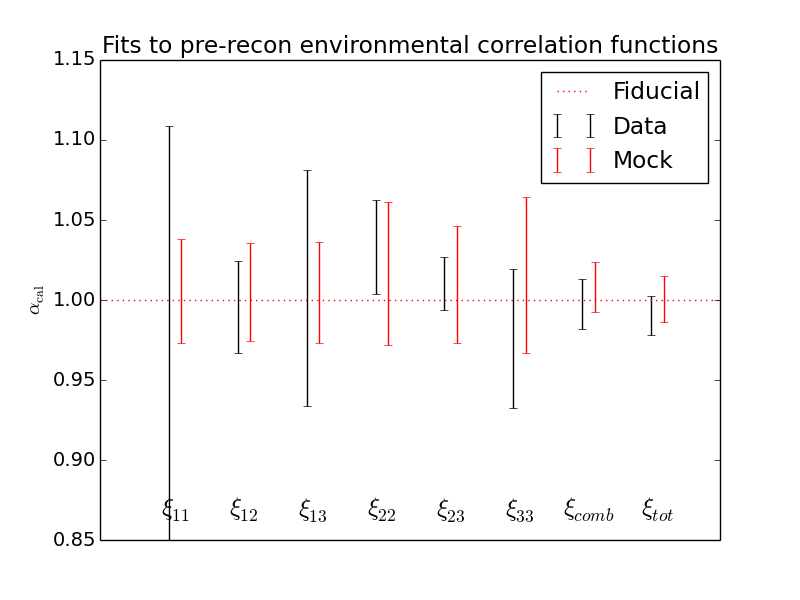}
\includegraphics[width=\columnwidth]{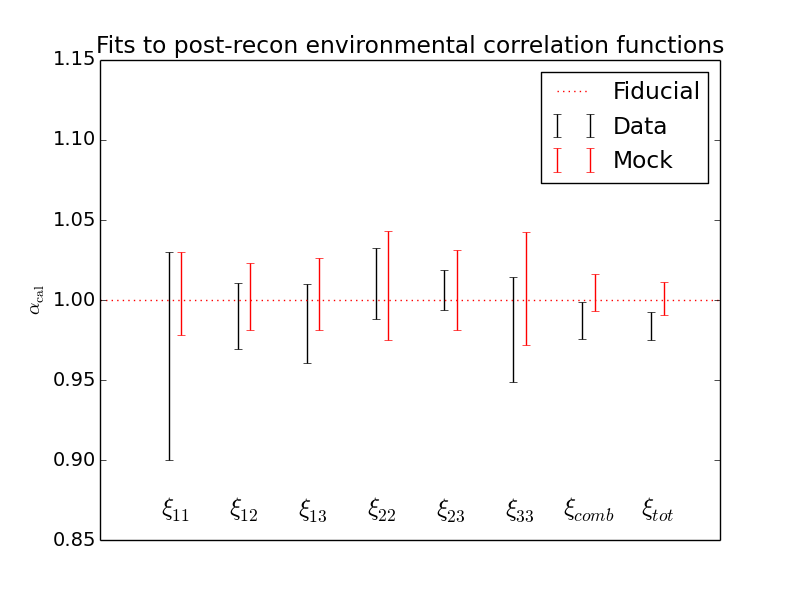}
\caption{The $68\%$ confidence range of the fits of $\alpha_{\rm cal}$
  to the pre-reconstruction (upper panel) and post-reconstruction
  (lower panel) auto-correlation and cross-correlation function
  measurements $\xi_{IJ}$ of the CMASS DR12 galaxy dataset, in three
  environmental slices defined by increasing local density.  The mean
  and standard deviation of the best-fitting values of $\alpha_{\rm
    cal}$ for the ensemble of QPM mock catalogues are also displayed.
  The last two pairs of results show the results of fitting a single
  $\alpha_{\rm cal}$ value to the set of measurements in different
  environments, including the appropriate covariance ($\xi_{\rm
    comb}$), and to the total galaxy correlation function ($\xi_{\rm
    tot}$).  These procedures produce similar final results.  The QPM
  mock mean template is used to construct the BAO model, although the
  results are not sensitive to this choice.}
\label{figalphafit}
\end{figure}

Figure \ref{figalphafit} compares the values of $\alpha_{\rm cal}$
obtained from each pre-reconstruction and post-reconstruction
environmental correlation function, displaying the $68\%$ confidence
ranges of the posterior probability distribution for the fit to the
data, and the mean and standard deviation of the best-fitting values
of $\alpha_{\rm cal}$ across the mock realizations.  The values of
$\alpha_{\rm cal}$ obtained from different environments are
consistent.  Fitting a single value of $\alpha_{\rm cal}$ to the six
individual post-reconstruction measurements (pre-reconstruction
results are given in brackets), using a covariance matrix of
$\alpha_{\rm cal}$ values deduced from the QPM mock catalogues, we
find $\alpha_{\rm cal} = 0.987 \pm 0.011$ ($0.997 \pm 0.016$) with a
minimum $\chi^2 = 1.6$ ($1.4$) for 5 degrees of freedom (six
$\alpha_{\rm cal}$ measurements minus one fitted parameter).  These
results are based on the BAO model constructed from the QPM mock mean
correlation function in each environment, although the same conclusion
holds when using the same CAMB model template to fit each environment.

The measurement of $\alpha_{\rm cal}$, produced by combining the
values in each environment, is consistent with the fit to the total
galaxy correlation function, $\alpha_{\rm cal} = 0.984 \pm 0.009$
($0.990 \pm 0.012$), albeit with a slightly inflated error.  We
attribute this increased error to the fact that, although the acoustic
peaks in each environmental correlation functions are detectable, they
are measured with a poorer statistical significance than for the total
correlation function.  The error in the distance-scale fit is a
sharply decreasing function of BAO detection significance in the
regime where the acoustic peak is just being resolved, changing faster
than the naive $\sqrt{\rm Volume}$ scaling we may associate with
sub-dividing a dataset.  Future galaxy redshift surveys will allow
acoustic peak detections in different data subsets in the
high-significance regime, which may allow an improvement in the BAO
fitting from sub-division into environments to be realized.

\subsection{Variation of the error in the BAO scale with environmental weighting}

Finally, we explored the effect on acoustic peak fits of assigning a
varying weight with environment when constructing the total galaxy
correlation function from the pair counts.  This weighting is
accomplished by the factor $x$ defined by Equation \ref{eqweight},
where $x=-1$ ($+1$) corresponds to assigning double weight to the
lowest (highest) density environment, and zero weight to the highest
(lowest) density environment.  For each new value of $x$, we
re-determined the template correlation function (again using the QPM
mock mean) and covariance matrix of the measurements by applying the
same procedure to the mock catalogues.  We note that this analysis
uses the original pair counts measured in 12 narrow density bins, and
does not use the 3 environmental correlation functions analyzed in the
previous section.

\begin{figure}
\includegraphics[width=\columnwidth]{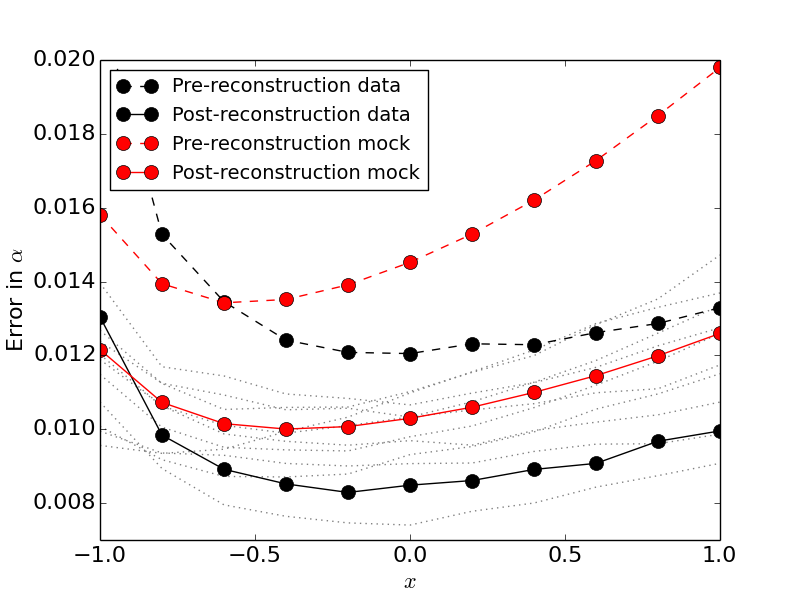}
\caption{The $68\%$ error in the distortion scale $\alpha_{\rm cal}$
  fitted to the data, or the standard deviation of the best-fitting
  $\alpha_{\rm cal}$ values to the ensemble of mocks, as a function of
  the parameter $x$ which controls the relative weighting of different
  environments to the pair count via Equation \ref{eqweight}.  Results
  are shown both pre- and post-reconstruction.  In order to indicate
  the effect of sample variance, results are also plotted for the
  first 10 mocks as the grey dotted lines.}
\label{figalphawei}
\end{figure}

Figure \ref{figalphawei} presents the $68\%$ error in the fitted
distortion scale $\alpha_{\rm cal}$ for the data, and the standard
deviation of the best-fitting $\alpha_{\rm cal}$ values for the
ensemble of mocks, as a function of the weighting parameter $x$, for
the pre-reconstruction and post-reconstruction correlation functions.
Using the ensemble of mocks, we find that moderately upweighting
underdense environments using $x \approx -0.5$ improves the standard
deviation in $\alpha_{\rm cal}$ by $8\%$ (pre-reconstruction, with the
scatter improving from $1.45\%$ to $1.34\%$) and $3\%$
(post-reconstruction, $1.03\%$ to $1.00\%$).  These results are
consistent with the notion that the acoustic peak is somewhat sharper
in underdense environments.  If the weight of underdense environments
is further increased, then the Poisson noise in the correlation
function starts to dominate, resulting in poorer measurements.

We find that weighting environments does not produce such a strong
enhancement in the accuracy of the standard ruler when applied to the
data, although a slight ($2\%$) improvement in the error is obtained
in the case of the post-reconstruction correlation function for $x =
-0.2$, slightly upweighting underdense environments.  These results
are consistent with the range of behaviours observed in different
realizations of the QPM mocks.  As an illustration of the sample
variance, in Figure \ref{figalphawei} we overlay the corresponding
trends for each of the first 10 mocks in the post-reconstruction case.

Our conclusions are qualitatively consistent with those reported in a
N-body simulation study by \citet{Achitouv15}, who also found that
upweighting underdense regions improves the accuracy of baryon
acoustic peak fits by a few per cent for a weighting parameter $x
\approx -0.5$.  Unlike \citet{Achitouv15}, our constraints in this
study degrade for $x < -0.5$, due to the increased Poisson noise of
the CMASS sample.  With a higher-density sample, \citet{Achitouv15}
were able to adopt a smaller smoothing scale more comparable to the
halo Lagrangian radii, and apply higher-order corrections to the
Zeldovich approximation.

\section{Conclusions}
\label{secconc}

The baryon acoustic peak, a robust prediction of early-universe
physics, is distorted by the growth of structure in the late-time
universe, imprinting additional cosmological information on the
feature.  We have measured the large-scale clustering properties of
the BOSS DR12 CMASS galaxy sample and corresponding mock catalogues,
within and between environmental slices defined by the local galaxy
overdensity $\delta$ smoothed on $15 \, h^{-1}$Mpc scales.  Our goal
was to delineate the dependence of the baryon acoustic peak scale and
shape on environment, and to test if weighting galaxies as a function
of environment improved the accuracy of the extracted distance scale.
Such enhancements may be possible if the acoustic peak template used
in model-fitting, or the accuracy of the displacements inferred by
density-field reconstruction, depend on environment.

The CMASS dataset permitted the measurement of baryon acoustic peaks
in each of the six auto-correlation and cross-correlation functions of
galaxies in three density slices $-1 < \delta < 0.01$, $0.01 < \delta
< 0.14$ and $\delta > 0.14$, where these divisions split the survey
into three sub-samples covering volumes in the ratio 7:2:3.  Given
that a linear power spectrum model may not provide a good description
of fluctuations within a slice of environments, we performed acoustic
peak fitting using two additional templates constructed as the mean of
the corresponding environmental correlations of two mock catalogues:
the QPM CMASS mocks, and dark matter N-body simulations constructed
from initial conditions including and excluding baryon acoustic
oscillations.  The mock mean correlation functions reveal that the
acoustic peak shape depends on environment, both before and after
density-field reconstruction, although the peak scale does not.

The standard-ruler scales fitted to the correlation functions assuming
these different templates are in close agreement, and the best-fitting
distances are consistent between the environments.  Fitting a single
distance scale across all the environments, with appropriate
covariance, we obtain a combined fit which is close to the result of
fitting the total correlation function, albeit with a $\sim 20\%$
greater distance error.  We attribute the somewhat larger error to the
reduced significance of the acoustic peak detections for individual
environments.

Assigning galaxies in underdense environments moderately higher
weights when measuring the total correlation function of the sample
lowers the scatter in the best-fitting distance scales by a few per
cent for the ensemble of QPM mocks, both before and after
density-field reconstruction.  Specifically, by up-weighting
underdense regions and down-weighting overdense regions by up to
$50\%$, the scatter in the preferred-scale fits to the ensemble of
mocks improves from $1.45\%$ to $1.34\%$ (pre-reconstruction) and
$1.03\%$ to $1.00\%$ (post-reconstruction).  This finding is
consistent with the broadening of the acoustic peak and reduced
accuracy of reconstruction in overdense environments.  If the weight
of underdense environments is further increased, then the Poisson
noise in the correlation function starts to dominate, resulting in
poorer measurements.  The gains in applying such weights to the DR12
CMASS dataset are not as evident as for the mocks, although the
observed trends are consistent with sample variance across the
ensemble of mocks.

As prospects for future work, studying perturbation theory models for
the environmental correlation function on BAO scales would allow the
construction of more accurate theoretical templates for BAO fitting to
environmental correlation functions, and for the extraction of the
other cosmological information encoded in the variation of clustering
with environment.  Another extension of this work would be to derive
the optimal environmental weight, considering both the variation of
the acoustic peak shape with environment and Poisson noise.  Finally,
future large-volume galaxy samples spanning a wider range of
environments than Luminous Red Galaxies, such as the Taipan Galaxy
Survey \citep{Cunha17}, may allow more significant improvements from
environmental weighting.

\section*{Acknowledgements}

We thank the anonymous referee for useful comments, and Cullan Howlett
for valuable input on a draft of this paper.  IA acknowledges funding
from the European Research Council under the European Community
Seventh Framework Programme (FP7/2007-2013 Grant Agreement no. 279954)
RC-StG EDECS.  Parts of this research were conducted by the Australian
Research Council Centre of Excellence for All-sky Astrophysics
(CAASTRO), through project number CE110001020.  Part of this work was
performed on the swinSTAR supercomputer at Swinburne University of
Technology.  This work was granted access to HPC resources of TGCC
through allocations made by GENCI, and we acknowledge support from the
DIM ACAV of the Region Ile de France.  We have used {\tt matplotlib}
\citep{Hunter07} for the generation of scientific plots, and this
research also made use of {\tt astropy}, a community-developed core
Python package for Astronomy \citep{Astropy13}.

\bibliographystyle{mnras}
\bibliography{baoenvpaper}

\bsp
\label{lastpage}
\end{document}